\documentstyle[subeqn]{article}
\begin{document}
\baselineskip = 18 pt

\begin{titlepage}
\begin{center}
\large {INVERSE SCATTERING METHOD AND VECTOR HIGHER ORDER NONLINEAR 
SCHR{\" O}DINGER EQUATION}
\end{center}
\vspace{2 cm}
\begin{center}
{\bf Sasanka Ghosh}$^*$ and {\bf Sudipta Nandy}$^{\dag}$\\
{\it Physics Department, Indian Institute of Technology, Guwahati}\\
{\it Panbazar, Guwahati - 781 001, India}\\
\end{center}
\vspace{3 cm}
\begin{abstract}
A generalized inverse scattering method has been developed for arbitrary $n$ 
dimensional Lax 
equations. Subsequently, the method has been used to obtain $N$ soliton 
solutions
of a vector higher order nonlinear Schr{\" o}dinger equation, proposed by us. It
has been shown that under suitable reduction, vector higher order nonlinear
Schr{\" o}dinger equation reduces to higher order nonlinear Schr{\" o}dinger 
equation. The infinite number of conserved quantities have been obtained by 
solving a set of coupled Riccati equations. A gauge equivalence is shown between
the vector higher order nonlinear Schr{\" o}dinger equation and the generalized
Landau Lifshitz equation and the Lax pair for the latter equation has also been 
constructed in terms of the spin field, establishing direct integrability of the
spin system.
\end{abstract}
\end{titlepage}
\newpage
\baselineskip 24 pt

\section{\bf Introduction}

Integrable models, since its inception, occupy a distinguised position because 
of their striking predictability and the existence of a special class of extended 
solutions, called solitons. Solitons are characterised by non-dispersive 
localised wave pulses which produce as a result of the competetion between  
linear dispersion and nonlinearity. Moreover, solitons retain thier shapes 
even after collisions among themselves and consequently exhibit particle like 
behaviour. Integrable models as well as solitons have applications in such 
diverse areas of physics as high energy physics, condensed matter physics, plasma
physics, nonlinear optics and nuclear physics. 

At present, a handful of nonlinear dynamical equations exist exhibiting soliton 
solutions. Historically, KdV equation \cite {1} is the most important one. It 
describes 
the dynamics of waves moving in shallow water. KdV and KP hierarchies also play
an important role in describing non-critical strings \cite {2} through the 
construction of $\tau$ functions. In KdV or KP hierarchy approach, evaluation of 
partition function and all correlation functions  may be made directly through
a set of integrable differential equations, bypassing the complicated 
technicalities involved in other approaches. Sine-Gordon equation \cite {3} is 
another 
welknown example of integrable equation, which first arose as a master equation 
for pseudo-spherical
surfaces. Besides, it describes the propagation of magnetic flux in a Joseption 
junction transmission line, the propagation of dislocations in a crystal lattice 
and many other physical problems. Nonlinear Schr{\" o}dinger equation and its 
higher order generalizations [4,5] have also 
many applications in plasma physics and in nonlinear optics. In particular recently, a lot of 
excitement is veered around the soliton solutions of a higher order 
generalization of nonlinear Schr{\" o}dinger equation, because of its potential
application in high speed fibre communication systems. The key observation is 
that the dynamics of the 
higher order nonlinear Schr{\" o}dinger equation (HNLS) \cite {6} not only takes 
care of 
dispersion loss, but also takes care of the propagation loss as the optical 
pulse propagates along the fibre. This is due to the fact that stimulated Raman 
effect \cite {6a}, which compensates the propagation loss, already exists withing 
the spectrum of HNLS equation. As a consequence, a very short pulse can be 
propagetd over a long distance without distortion.  
Thus, it is optical solitons, which
are responsible behind the successful propagation of optical pulses through 
the fibre over a long distance. Therefore, the importance of the study of 
integrable models and to find their soliton solutions is unquestionable.  

There are several methods exist to obtain soliton solutions of a nonlinear 
evolution equation, like Hirota bilinear method \cite {7}, Painleve analysis 
\cite {8}, B{\" a}cklund transformation \cite {9} and inverse scattering 
transform (IST) \cite {10}. IST method, however, 
is the most elegant tool, which eventually proves the complete integrability of 
the evolution equation. IST method, on the other hand, is the most complicated 
one and is intimately connected with the existence of an auxiliary linear 
problem, known as Lax equations. Most of the known integrable evolution 
equations are, in fact, associated with $2\times 2$ and $3\times 3$ dimensional 
Lax operators. But 
there are some evolution equations whose associated linear equations, in general, 
cannot be cast in terms of $2\times 2$ or $3\times 3$ Lax operators. One such 
example is higher order nonlinear Schr{\" o}dinger equation (HNLS) \cite {6}:
\begin{equation} 
i\partial_t q + {\beta }_1 \partial_{xx} q + {\beta }_2 \vert q\vert^2 q 
i\epsilon [\beta_3 \partial_{xxx}q + \beta_4 \partial_x(\vert q\vert^2)q 
+ \beta_5 \vert q\vert^2\partial_x q] = 0
\label{1.1}
\end{equation}
where, $q$ is a complex field and $\beta_i\quad i=1,2\cdots,5$ are the constant 
coefficients. Notice that in absense of last three terms, (\ref {1.1}) reduces 
to nonlinear Schr{\" o}dinger equation. It is observed recently, for arbitrary 
values of the coefficients, (\ref {1.1}) is associated with $n\times n$ Lax 
operators \cite {11}. Interestingly, the parameters, $\beta$ are related to the 
dimension of the Lax operators. Unfortunately, IST method is not well understood 
for arbitrary dimensional Lax oparators. Nonetheless, IST methods 
developed in the context of $2\times 2$ \cite {10} and 
$3\times 3$ \cite {12,12a} matrix Lax operators cannot be generalized straight
forwardly to handle more 
general cases. Thus, in order to obtain soliton solutions of (\ref {1.1}) by 
IST method, a generalized IST method, associated with $n$ dimensional Lax 
operator, is to be formulated. In this paper, we will, therefore, develop a 
generalized IST method, which ultimately leads to solving a set of coupled 
Gelfand Levitan Marchenko equations \cite {12b} for obtaining soliton solutions.  
We will use the generalized IST method to obtain soliton solutions for an 
evolution equation, whose dynamical field is 
an $(n-1)$ component vector, $\vec q$. $\vec q$, in fact, is a vector 
generalization of the field $q$ given in (\ref {1.1}). 
The evolution equation may, therefore, be called vector higher
order nonlinear Schr{\" o}dinger equation (VHNLS) and will be described 
in the next section. We will also show that under suitable reduction
VHNLS equation reduces to (\ref {1.1}). 

The rigidity in the structures of the solitons reveals that the system has a 
huge underlying symmetry. This symmetry is, in general, manifested by the 
existence of an infinite number of conserved quantities. Existence of infinite 
number of conserved quantities
is a key criterion for a field theoretical model to be called integrable in 
Liouville sense \cite {13}. Construction of Lax pair of a dynamical system 
although itself is a good hint for integrability, many integrable models may not 
satisfy Liouville integrability criterion. In this paper we will obtain Riccati 
equations associated with an 
$n$ dimensional Lax oparator and indentify the generating function of the 
conserved charges. Consequently, we will find conserved charges explicitly 
in terms of the field variables and their derivatives.  

Another interesting aspect, we will address, is to establish a connection 
between nonlinear field theories and spin systems. In this context, it is 
welknown that 
nonlinear Schr{\" o}dinger equation is gauge related to Landau Lifshitz equation 
\cite {14}. Recently Sasa Satsuma equation \cite {14a}, a particaluar version of HNLS 
equation is shown to be gauge related to a generalized Landau Lifshitz equation, 
where the spin field is associated with $SU(3)$ group \cite {15,16}. It is thus 
expected that a spin system may also exist corresponding to a vector 
generalisation of HNLS equation. We will
establish a connection between the VHNLS equation, proposed by us and a 
generalized Landau Lifshitz equation and obtain a Lax pair for the spin system.   

The organization of the paper is as follows. In section 2, the VHNLS equation 
is introduced and a Lax pair for the nonlinear equation is  
constructed. We also show a reduction procedure under which VHNLS equation 
reduces to a two parameter family of HNLS equation. In section 3, we show the 
existence of infinite number of conserved quantities, by solving a set of 
coupled 
Riccati equations. We develop the inverse scattering method for an $n$ 
dimensional Lax operators in section 4 and obtain $n$ Gelfand Levitan Marchenko 
equations. Section 5 deals with the solutions  ofGelfand Levitan Marchenko equations 
and cosequently find $N$ soliton solutions. We show in section 6 the gauge 
equivalence of VHNLS equation and the generalised Ladau Lifsitz equation and 
obtain a Lax pair for the generalized Landau Lifshitz equation in terms of the 
spin field. Section 7 is the concluding one.

\section{\bf Evolution Equation and Lax Pair}

We propose a vector nonlinear evolution equation of the form
\begin{equation}
\vec{q}_t + \epsilon \vec{q}_{xxx} + 3\epsilon (\vec{q^*}\cdot\vec{q}_x) \vec{q}
+ 3\epsilon \vert \vec{q}\vert^2 \vec{q}_x = 0
\label{2.1}
\end{equation}
where, the dynamical field variable, $\vec{q} = (q_1,q_2,\cdots ,q_{n-1})$, is 
an $(n-1)$-tuple vector. The equation (\ref {2.1}) is an example of vector 
higher order nonlinear Schr{\" o}dinger equation (VHNLS). The vector field 
$\vec q$ may be interpreted as an $(n-1)$ interacting optical modes describing 
the dynamics of a charge field with $(n-1)$ colours. In fact, we will show 
under suitable reduction, (\ref {2.1}) yields to the HNLS equation \cite {11}. 

In order to obtain Lax pair for (\ref {2.1}), we first introduce the $n\times n$ 
matrix linear eigenvalue problem as
\begin{subequations}
\begin{eqnarray}
\partial_x \Psi &=& {\bf U}(x,t,\lambda)\Psi \\     
\label{2.2a}
\partial_t \Psi &=& {\bf V}(x,t,\lambda)\Psi
\label{2.2b}
\end{eqnarray}
\label{2.2}
\end{subequations}
where, ${\bf \Psi }(x,t)$ is an $n$-tuple vector auxilliary field, $\lambda $ 
is the spectral parameter and the Lax 
operators ${\bf U}(x,t)$ and ${\bf V}(x,t)$ are $n\times n$ matrices. Let us 
now assume an explicit form of the Lax pair, ${\bf U}(x,t)$ and 
${\bf V}(x,t)$, associated with the VHNLS equation, as  
\begin{subequations}
\begin{eqnarray}
{\bf U} &=& -i \lambda {\bf \Sigma} + {\bf A} \\
{\bf V} &=& - \epsilon {\bf A}_{xx} + \epsilon ({\bf A}_x{\bf A} 
          -{\bf A}{\bf A}_x) + 2\epsilon {\bf A}^3 \\ \nonumber
     & &   -2i\epsilon \lambda {\bf \Sigma} ({\bf A}^2 -{\bf A}_x) 
        + 4\epsilon {\lambda}^2{\bf A}
        -4i\epsilon {\lambda}^3{\bf \Sigma},
\end{eqnarray}
\label{2.3}
\end{subequations}
In the equation (\ref {2.3}) 
${\bf \Sigma }$ is a c-no. diagonal matrix and the matrix ${\bf A}(x,t)$ 
consists of dynamical fields, $\vec{q}(x,t)$ and $\vec{q^*}(x,t)$ only. It is
interesting to note that the evolution equation for the matrix ${\bf A}(x,t)$ 
immediately follows from the compatibility condition, namely 
\[ 
{\bf U}_t(x,t) - {\bf V}_x(x,t) + [{\bf U}(x,t) , {\bf V}(x,t)] = 0 
\]
provided ${\bf A}$ and ${\bf \Sigma}$ satisfy the conditions that
\begin{equation}
{\bf \Sigma}^2 = {\bf 1}, \quad\quad 
{\bf \Sigma}{\bf A} + {\bf A}{\bf \Sigma} = 0
\label{2.4}
\end{equation}
and as a consequence, the nonlinear evolution equation for ${\bf A}(x,t)$ 
becomes
\begin{equation}
{\bf A}_t + \epsilon {\bf A}_{xxx} - 3\epsilon ({\bf A}^2 
{\bf A}_x + {\bf A}_x {\bf A}^2) = 0
\label{2.5}
\end{equation}
It is now clear that various representations of the matrix ${\bf A }$ 
in terms of the fields $\vec{q}$ and $\vec{q^*}$ yield to different nonlinear 
evolution equations.  
To associate (\ref {2.5}) with the VHNLS equation (\ref {2.1}), let us consider 
the explicit expressions of ${\bf \Sigma }$ and ${\bf A}(x,t)$ of the form 
satisfying the properties (\ref {2.4}) as 
\begin{subequations}
\begin{eqnarray}
{\bf \Sigma} &=& \sum_{i=1}^{n-1} e_{ii} - e_{nn} \\ 
\label{2.6a}
{\bf A}(x,t) &=& \sum_{i=1}^{n-1} q_i(x,t) e_{in}
 - \sum_{i=1}^{n-1} q_i^*(x,t) e_{ni} 
\label{2.6b}
\end{eqnarray}
\label{2.6}
\end{subequations}
where, $e_{ij}$ is an $n\times n$ matrix whose only $(ij)$th element is unity, 
the rest elements being zero and $q_i(x,t)$ is the $i$th. component of the 
dynamical field $\vec q$. Substituting (\ref {2.6}) in (\ref {2.5}), the
evolution equation becomes
\begin{equation}
q_{it} + \epsilon q_{ixxx} + 3\epsilon (\sum_{j=0}^{n-1} q_j^*.q_{jx}) q_i
+ 3\epsilon (\sum_{j=0}^{n-1} q_j^*q_j) q_{ix} = 0
\label{2.7}
\end{equation}
which is nothing but VHNLS equation (\ref {2.1}), written in the component form. 
Now in order 
to obtain HNLS equation (\ref {1.1}) let us consider the following reduction. Notice that
all the dynamical fields $q_i$ in (\ref {2.7}) are independent. However, instead 
of $(n-1)$ independent dynamical fields, if we restrict ourselves to only one 
dynamical field $q$ and its complex conjugate $q^*$ in (\ref {2.7}), then all 
the $q_i$s are not independent. $q_i$s, in this case, may be chosen as either  
$q$ or $q^*$. 
If we, for example, choose $m$ number of $q_i$ as $q$ and the rest $(n-m-1)$ 
numbers as $q^*$, (\ref {2.7}) reduces to HNLS equation  
\begin{equation}  
q_t + \epsilon q_{xxx} + 6m\epsilon\vert q\vert^2 q_x 
+ 3(n-m-1)\epsilon (\vert q\vert^2)_x q = 0.
\label{2.8} 
\end{equation}
The equation (\ref {2.8}), to be precise, is a gauge equivalent version of HNLS 
equation. Soliton solutions of (\ref {2.8}) and those of HNLS equation are, in 
fact, related 
through a U(1) gauge tranformation \cite {11}. Two well known euqations 
immediately 
follow from (\ref {2.8}). If we choose $m=n-1$, (\ref {2.8}) gives rise to 
Hirota equation \cite {17} 
\begin{equation}  
q_t + \epsilon q_{xxx} + 6\epsilon\vert q\vert^2 q_x = 0 
\label{2.9} 
\end{equation}
after rescalling of $q$ and $q^*$ as $q=(n-1)^{-\frac{1}{2}}q$ and 
$q^*=(n-1)^{-\frac{1}{2}}q^*$ respectively. On the other hand, if we choose 
$m=(n-1)/2$, 
which is only possible for odd dimensional Lax pair, (\ref {2.8}) reduces to 
Sasa Satsuma euqation \cite {14a,16,18},
\begin{equation}  
q_t + \epsilon q_{xxx} + 6\epsilon\vert q\vert^2 q_x 
+ 3\epsilon (\vert q\vert^2)_x q = 0,
\label{2.10} 
\end{equation}
once again with appropriate scalling of the fields $q$ and $q^*$ respectively as 
$q=(\frac{n-1}{2})^{-\frac{1}{2}}q$ and $q^*=(\frac{n-1}{2})^{-\frac{1}{2}}q^*$. 
\vspace{.5 cm}

\section {\bf Riccati Equation and Conserved Charges}

In order to obtain Riccati equation, we first write the Lax equation 
(\ref {2.2}a,\ref {2.3}a,\ref {2.6}) in the 
component form. For the first $(n-1)$ components of ${\bf \Psi}$, (\ref {2.2}a) 
can be written in the form,
\begin{eqnarray*}
 \Psi_{1x} &=& -i\lambda \Psi_1 + q_1  \Psi_n\\ 
 \Psi_{2x} &=& -i\lambda \Psi_2+ q_2  \Psi_n\\ 
 & &\vdots\\
 \Psi_{n-1x} &=& -i\lambda \Psi_{n-1} + q_{n-1} \Psi_n\\
\end{eqnarray*}
{\it i.e.}  
\begin{subequations}
\begin{equation}
 \Psi_{ix} = -i\lambda \Psi_i + q_i  \Psi_n 
\label{3.1}
\end{equation}
where, $i = 1,2,\cdots , n-1 $ and $\Psi_i$ denotes the $i$th. component of 
${\bf \Psi}$. But for the $n$th. component of ${\bf \Psi}$, the Lax equation 
(\ref {2.2}a) has a different form as 
\begin{eqnarray}
 \Psi_{nx} &=& i\lambda \Psi_n - q_1^* \Psi_1  
 - q_2^* \Psi_2  \cdots - q_{n-1}^* \Psi_{n-1}\nonumber \\
&=& i\lambda\Psi_n - \sum_{j=1}^{n-1}q_j^* \Psi_j
\label{3.2}
\end{eqnarray}  
\end{subequations}
Following now a similar procedure as in \cite {16} we write,
\begin{equation}
\Gamma_i = \frac{\Psi_i}{\Psi_n},
\label{3.3}
\end{equation}
$i = 1,2,\cdots, n-1$, which are related to the conserved charges 
$\alpha_{nn}(\lambda)$ in the following way,
\begin{eqnarray}
\ln \alpha_{nn}(\lambda) &=& 
\ln \Psi_n -i\lambda x\vert_{x\rightarrow \infty}\nonumber \\
                         &=& -\int_{-\infty}^{\infty}dx
                            (\sum_{i=1}^{n-1}q_i^*\Gamma_i)
\label{3.4}
\end{eqnarray}
We will see in section 5 that $\alpha_{nn}(\lambda)$ is, indeed, $(nn)$th
element of the scattering data matrix and more so it does not evolve with 
time. By using (\ref{3.1},b) and (\ref{3.3}), we may obtain a first order
differential equation for each $\Gamma_i$, 
\begin{equation}
\Gamma_{ix} + 2i\lambda\Gamma_i - \sum_{j=1}^{n-1}q_j^*\Gamma_j\Gamma_i
     - q_i = 0\\
\label{3.5}
\end{equation}
The set of $(n-1)$ coupled nonlinear differential equations for $\Gamma_i$ in 
(\ref {3.5}) are called 
Riccati equations. It is obvious from (\ref {3.4}) that the solutions of Riccati 
equations eventually determine the conserved quantities. Now in order to solve 
(\ref{3.5}), we assume a series solution of $\Gamma_i$ as 
\begin{equation}
\Gamma_i(x,\lambda) = \sum_{n=0}^{\infty}C^i_n(x)\lambda^{-n} 
\label{3.6}
\end{equation}
Substituting (\ref{3.6}) into (\ref{3.5}), the following recursion relations may 
be obtained.
\begin{subequations}
\begin{eqnarray}
C_0^i = 0 ;\quad \quad \quad  C^i_1 = \frac{q_i}{2i}
\end{eqnarray}
and
\begin{equation}
2iC_{k+2}^i + (C^i_{k+1})_x 
- \sum_{m=0}^{k-1}C^i_{k-1+m}\sum_{j=1}^{n-1}q_j^*C^j_m =0\\
\label{3.7}
\end{equation}
\end{subequations}
with $ k= 0,1,2,..........  $. The infinite number of Hamiltonians (conserved
quantities) may explicitly be determined in terms of the dynamical field 
variables $q_i$ and their derivatives by expanding $\alpha_{nn}(\lambda)$
in the form,
\begin{equation}
ln\alpha_{nn}(\lambda) = (n-1)\sum_{l=0}^{\infty}\frac{(-1)^l}{(2i)^{2l +1}}
H_l\lambda^{-l}
\label{3.8}
\end{equation}
and thus comparing (\ref {3.8}) with (\ref {3.4}) and (\ref {3.6}), ${H_l}$ 
becomes 
\begin{equation}
H_l = \frac{(2i)^{2l+1}}{(-1)^l(n-1)}\int dx[\sum_{j=1}^{n-1}q^*_jC^j_l)]
\label{3.9}
\end{equation}
The explicit expresseions of the first few low order Hamiltonians are given 
below.
\begin{subequations}
\begin{eqnarray}
H_1 &=& \frac{1}{n-1}\int dx [\sum^{n-1}_{j=1}q_j^*q_j] \\
\label{3.10a}
H_2 &=& \frac{1}{2(n-1)}\int dx \sum_{j=1}^{n-1}[q_j^*q_{jx} 
- q^*_{jx}q_j]\\
\label{3.10b}
H_3 &=& \frac{1}{n-1}\int dx [(\sum_{j=1}^{n-1}q_j^*q_j)^2 - \sum_{j=1}^{n-1}
q_{jx}^*q_{jx}]\\
\label{3.10c}
H_4 &=& \frac{1}{2(n-1)}\int~dx\sum_{j=1}^{n-1}[q_j^*q_{jxxx}-q^*_{jxxx}+ 
3(\sum_{k=1}^{n-1}\vert q_k\vert)^2(q_j^*q_{jx}-q^*_{jx}q_j)]\nonumber\\
&~&\\
\label{3.10d}
H_5 &=& \frac{1}{n-1}\int dx[\sum_{j=1}^{n-1}q_{jxx}^*q_{jxx} 
+ 2(\sum_{j=1}^{n-1}q^*_jq_j)^3 - (\sum_{j=1}^{n-1}(\vert q_j\vert^2)_x)^2 
\nonumber \\
&-& 4\sum_{k=1}^{n-1}\vert q_j\vert^2\sum_{j=1}^{n-1}q_{jx}^*q_{jx} 
- 2\sum_{j=1}^{n-1}q_j^*q_{jx}\sum_{k=1}^{n-1}q^*_{kx}q_k] 
\label{3.10e}
\end{eqnarray}
\label{3.10}
\end{subequations}
We verify directly by using equations of motions that $H_1$, $H_2$, $H_3$, 
$H_4$ and $H_5$ are, indeed, constants of motions. 

Let us now specialize to HNLS equation. If we assume that $m$ number of $q_i$s 
are chosen as $q$ and the rest $n-m-1$ as $q^*$, the Hamiltonians in 
(\ref {3.10}) reduce to
\begin{subequations}
\begin{eqnarray}
H_1 &=& \int dx \vert q \vert^2\\
\label{3.11a}
H_2 &=& (2m -n+1)\int dx (q_xq^* - qq^*_x)\\ 
\label{3.11b}
H_3 &=& \int dx ((n-1)\vert q \vert^4 - q_x q^*_x)\\
\label{3.11c}
H_4&=&(2m-n+1)\int~dx [3(n-1)\vert q\vert^2 (q_xq^*-qq^*_x) 
+\frac{1}{2}(q_{xxx}q^*-qq^*_{xxx})]\nonumber \\ 
&~&\\
\label{3.11d}
H_5&=&\int~dx [q_{xx}q^*_{xx}+2(n-1)^2 \vert q \vert^6
      +[(2m-n-1)^2 -4(n-1)]\vert q \vert^2 q_x q^*_x\nonumber \\ 
      &-&[(n-1) + \frac{2m(n-l-1)}{(n-1)}]((\vert q \vert^2)_x)^2]\\ \nonumber  
&~&
\label{3.11e}
\end{eqnarray}
\label{3.11}
\end{subequations}

Notice that $H_1$ in (\ref {3.11}a) has a universal form, which implies that all
solitons have the same energy irrespective of their shapes. However, the 
momentum, $H_2$ crucially depends  on the numbers of $q$, chosen in a given 
representation of the matrix ${\bf A}$ (\ref {2.6}). For example, $H_2$ becomes
zero for Sasa Satsuma case, {\it i.e.} for $m=(n-1)/2$. In fact, all  
conserved quantities having even indices
become trivial for Sasa Satsuma case. $H_3$ in (\ref {3.11}c) also has a 
somewhat universal form, depending only on the dimensions of the Lax pairs. But 
the higher order conserved quantities starting from $H_5$ do not possess such
universal forms. Their explicit forms depend both on the dimensionality of the 
matrix Lax pair and also on the representations of the matrix ${\bf A}$.

\section{\bf Generalized Gelfand Levitan Marchenko Equations}

We now generalize IST method suitable for studying $n$ dsmensional Lax 
operators.  The first 
step in this direction is to obtain a set of generalized Gelfand Levitan 
Marchenko equations. This generalisation is a nontrivial one and crucially 
depends on the properties of scattering data matrix. However, we will see that 
for three dimenisonal Lax oparators generalized Gelafand Levitan Marchenko 
equations will reduce to Sasa Satsuma case \cite {14a} as is expected. We broadly follow 
the 
treatment of Manakov, developed in the context of $3\times 3$ Lax operator      \cite {12a} and 
for that assume Jost functions, for real $\lambda$, satisfy the boundary 
conditions,
\begin{subequations}
\begin{equation}
\Phi^{(i)} = e_i e^{-i\lambda x}
\label{IV1a} 
\end{equation} 
with $ i=1,2.......n-1$, but the $n$th. one satisfies a different boundary 
condition like
\begin{equation}
\Phi^{(n)} = e_n e^{i\lambda x}  
\label{IV1b}
\end{equation}
\label{IV1}
\end{subequations}
as $x\rightarrow -\infty$. Similarly, as $x\rightarrow \infty$, other set of 
Jost functions satify the boundary conditions
\begin{subequations}
\begin{equation}
\Psi^{(i)} = e_i e^{-i\lambda x}
\label{IV2a}
\end{equation}
 for $ i= 1,.......n-1$ and for the $n$th. one,
\begin{equation}
\Psi^{(n)} = e_n e^{i\lambda x}.
\label{IV2b}
\end{equation}
\label{IV2}
\end{subequations}
In the equations (\ref {IV1}) and (\ref {IV2}) $e_i$s are the basis vectors for 
an n-dimensional vector space. It follows from (\ref {2.6}b) that 
${\bf A}^{\dagger}=-{\bf A}$ for real valued $\lambda$ and thus we have
\begin{equation}
\partial_x({{\bf \Psi}^{(1)}}^{\dagger}{\bf \Psi}^{(2)})= 0
\label{IV3}
\end{equation}
for any pair of solutions of equation (\ref {2.1}a), ${\bf \Psi}^{(1)}$ and 
${\bf \Psi}^{(2)}$, having the same eigenvalue. It is straightforward to show from 
(\ref{IV1}) and (\ref{IV2}) that 
\begin{equation}
\Phi^{(i)\dagger}\Phi^{(j)} = \Psi^{(i)\dagger}\Psi^{(j)}=\delta_{ij} 
\label{IV4}
\end{equation}
for $i,j =1,2,.........n$.
Since the set of Jost functions ${\bf \Psi_i}$ are linearly independent and 
the maximum number of independent Jost functions is $n$, we may 
express the set of Jost functions $\Phi_i$ as a linear combination of 
$\Psi_i$ as 
\begin{equation}
\Phi^{(i)}(x,\lambda) = 
\sum_{j=1}^n \alpha_{ij}(\lambda)\Psi^{(j)}(x,\lambda)
\label{IV5} 
\end{equation}
where $\alpha_{ij}(\lambda)$ is the $(ij)$th element of scattering data matrix, 
which can be expressed by using (\ref{IV4}) and (\ref {IV5}) in the form
\begin{equation}
\alpha_{ij}(\lambda)= \Psi^{(j)\dagger}(x,\lambda)\Phi^{(i)}(x,\lambda)
\label{IV6}
\end{equation}
The orthogonality property of the scattering data matrix elements for real 
eigenvaule $\lambda$, subsequently follows from (\ref{IV1}), (\ref{IV2}) and 
(\ref{IV6}) and thus we obtain
\begin{equation}
\sum_{k=1}^{n}\alpha_{ik}(\lambda)\alpha_{jk}(\lambda) =\delta_{ij}
\label{IV7}
\end{equation}     
which finally gives
\begin{equation}
\Psi^{(i)}(x,\lambda) =
 \sum_{j=1}^n \alpha^*_{ji}(\lambda)\Phi^{(j)}(x,\lambda)
\label{IV8}
\end{equation}

It is further interesting to see, by exploiting the properties of $[\alpha_{ij}]$ 
and $[\alpha^*_{ij}]$ matrices, that we can write the element $\alpha^*_{ij}$ as 
the cofactor of the elements of the matrix $[\alpha_{ij}]$. In particular, 
$\alpha^*_{ni}$ element can be written as 
\begin{equation}
\alpha_{ni}^* = (-1)^{n+i}det[\tilde{\alpha}_{ni}]
\label{IV9}
\end{equation} 
where $[\tilde{\alpha}_{ni}]$ is a $(n-1)\times (n-1)$ matrix, constructed from 
the $n\times n$ scattering matrix, $[\alpha_{ij}]$ with nth row and ith column 
being omitted, {\it i.e.} $det[\tilde{\alpha}_{ni}]$ is the minor of  
$\alpha_{ni}$ element of scattering matrix, $[\alpha_{ij}]$.  
Now by using (\ref{IV5}) and (\ref{IV9}), we obtain the following useful 
relations among the Jost functions $\Phi^{(i)}$ and $\Psi^{(i)}$. The first 
$n-1$ Jost functions in (\ref {IV5}) satisfy
\begin{subequations}
\begin{equation}
\frac{1}{\alpha^*_{nn}(\lambda)}\sum_{j=1}^{n-1}(Adj [\tilde{\alpha}_{nn}])_
{kj}\Phi^{(j)}e^{i\lambda x} = \Psi^{(k)}e^{i\lambda x} -
\frac{\alpha^*_{nk}}{\alpha^*_{nn}}\Psi^{(n)}e^{i\lambda x}
\label{IV10a}
\end{equation}
with $k=1,........n-1$, but the $n$th. Jost function,$\Phi^{(n)}$ obey the 
relation  
\begin{equation}
\frac{1}{\alpha_{nn}} \Phi^{(n)}e^{-i\lambda x} =
 \Psi^{(n)}e^{-i\lambda x} + \frac{1}{\alpha_{nn}} \sum_{j=1}^{n-1}
\alpha_{nj}\Psi^{(j)}e^{-i \lambda x}
\label{IV10b}
\end{equation}
\label{IV10}
\end{subequations}
Notice that in deriving (\ref{IV10}), we have used the following properties of 
the scattering data matrix. 
\[
\alpha^*_{nk}\delta_{ij} = \sum_{l=1}^{n-1}[\tilde{\alpha_{nk}}]_{il}
(Adj [\tilde{\alpha}_{nk}])_{lj}
\]
It is important to mention that analyticity properties of the  Jost functions 
and the elements of scattering matrix may be obtained from (\ref{IV10}). 

We are now going to derive the Gelfend, Levitan and Marchencho equation for an 
$n$ dimensional Lax pair. Let us consider an integral representation of the Jost 
function $\Psi^{(i)}$ for $(i=1,2,.....,n)$. For the first $n-1$ Jost functions 
we may choose the following integral representations 
\begin{subequations}
\begin{equation}
\Psi^{(j)}(x,\lambda) =e_j e^{-i\lambda x} + 
\int_x^{\infty}dy {\bf K}^{(j)}(x,y)e^{-i\lambda y}
\label{IV11a}
\end{equation}
with $j=1,2,.....n-1$, while the $n$th Jost function may be written as
\begin{equation}
\Psi^{(n)}(x,\lambda) =e_n e^{i\lambda x} + 
\int_x^{\infty}dy {\bf K}^{(n)}(x,y)e^{i\lambda y}.
\label{IV11b}
\end{equation} 
\label{IV11}
\end{subequations}
where, $e_i,\quad i=1,2,\cdots,n$ are basis vectors for an $n$ dimensional 
vector space and the kernels ${\bf K}^{(j)}$ and 
${\bf K}^{(n)}$ are $n$ dimensional column vectors, which may be written 
explicitly in the component form as
\begin{subequations}
\begin{equation}
{\bf K}^{(j)}(x,y)=\sum_{m=1}^n K_m^{(j)}(x,y)e_m
\label{IV12a}
\end{equation}
\begin{equation}
{\bf K}^{(n)}(x,y)=\sum_{m=1}^n K_m^{(n)}(x,y)e_m
\label{IV12b}
\end{equation}
\label{IV12}
\end{subequations} \newpage
Substituting (\ref{IV11}) in (\ref{IV10b}), we obtain
\begin{eqnarray}
\frac{1}{\alpha_{nn}}\Phi^{(n)} &=& e_n e^{i\lambda x} +\int_x^{\infty}
dy {\bf K}^{(n)}(x,y)e^{i\lambda y} + \sum_{j=1}^{n-1}
\frac{\alpha_{nj}}{\alpha_{nn}}e_j e^{-i \lambda x} \nonumber \\
&+& \sum_{j=1}^{n-1}\frac{\alpha_{nj}}{\alpha_{nn}}\int_x^{\infty}
dy {\bf K}^{(j)}(x,y)e^{-i\lambda y} 
\label{IV13}
\end{eqnarray}
Multiplying now both sides of (34) by $\frac{1}{2\pi}
\int_{-\infty}^{\infty}d\lambda e^{-i\lambda z}$, with an assumption  $z>x$
and using the analyticity property of the Jost function, it follows that
\begin{subequations}
\begin{eqnarray}
&~&\frac{1}{2\pi}\int_{-\infty}^{\infty} d\lambda \frac{\Phi^{(n)}}
{\alpha_{nn}(\lambda)}
e^{-i\lambda z} = {\bf K}^{(n)}(x,z) + \int_{-\infty}^{\infty}\frac
{d\lambda}{2\pi}\sum_{j=1}^{n-1}\frac{\alpha_{nj}(\lambda)}
{\alpha_{nn}(\lambda)}e_j e^{-i\lambda(x+z)} \nonumber \\
&+&\int_{-\infty}^{\infty}\frac{d\lambda}{2\pi}\int_{x}^{\infty}dy
\sum_{j=1}^{n-1}
\frac{\alpha_{nj}(\lambda)}{\alpha_{nn}(\lambda)}{\bf K}^{(j)}(x,y)
e^{-i\lambda(y+z)} \\ \nonumber
&~& 
\label{IV14a}
\end{eqnarray}
The L.H.S. of (\ref{IV14}a) can be simplified further by taking into account 
that $\frac{1}{\alpha_{nn}(\lambda)}$ is analytic in the lower half plane 
except at the points, say $\lambda_j^*$ with $Im \lambda_j^*>0$ and 
$j=1,2........N$, where $\frac{1}{\alpha_{nn}(\lambda)}$ has $N$ simple poles. 
Moreover, we assume that at the simple poles $\lambda_j^*$, $\Phi^{(n)}$ to be
of 
the form
\begin{equation}
\Phi^{(n)}(x,\lambda_j^*)=\sum_{p=1}^{n-1}C^{(j)}_{np}\Psi^{(p)}
(x,\lambda_j^*)
\label{IV14b}
\end{equation}
With these assumptions, L.H.S. of (\ref{IV14}a) becomes 
\begin{equation}
-i\sum_{j=1}^N\frac{e^{-i\lambda_j^*z}}{\alpha^{\prime}_{nn}(\lambda^*_j)}
\sum_{p=1}^{n-1}C_{np}^{(j)}\Psi^{(p)}(x,\lambda_j^*)
\label{IV14c}
\end{equation}
where $\alpha_{nn}^{\prime}$ denotes derivative with respect to $\lambda $. 
By Substituting the integral representations of $\Psi^{(p)}(x,\lambda_j^*)$ 
from (\ref {IV11a}), (\ref {IV14c}), {\it i.e.} the L.H.S. of (\ref {IV14}a) 
finally reduces to 
\begin{equation}
-i\sum_{j=1}^N \frac{e^{-i\lambda_j^*z}}{\alpha_{nn}^{\prime}(\lambda_j^*)}
\sum_{p=1}^{n-1}C_{np}^{(j)} [e_p e^{-i\lambda_j^* x} +\int_x^{\infty}dy
{\bf K}^{(p)}(x,y)e^{-i\lambda_j^* y}]
\label{IV14d}
\end{equation}
\label{IV14}
\end{subequations} 
Let us now introduce a funtion $F_p(x+y)$ as 
\begin{equation}
F_p(x+y)=i\sum_{j=1}^{N}\frac{C_{np}^{(j)}e^{-i\lambda^*_j(x+y)}}
{\alpha_{nn}^{\prime}(\lambda^*_j)} +
\int_{-\infty}^{\infty}\frac{d\lambda}{2\pi}\frac{\alpha_{np}(\lambda)}
{\alpha_{nn}(\lambda)}e^{-i\lambda(x+y)}
\label{IV15}
\end{equation}
In terms of the function, $F_p(x+y)$ in (\ref {IV15}), (\ref{IV14}a) and 
(\ref{IV14d}) together may be written in a compact form as
\begin{subequations}
\begin{equation}
{\bf K}^{(n)}(x,z) + \sum_{p=1}^{n-1}e_pF_p(x+z) + \sum_{p=1}^{n-1}
\int_x^{\infty}{\bf K}^{(p)}(x,y)F_p(z+y) =0
\label{IV16a}
\end{equation}
The inetgral equation (\ref {IV16a}) is one of the desired Gelfand Levitan 
Marchenko 
equations for the kernel ${\bf K}^{(n)}$. Other integral equations for the 
kernels ${\bf K}^{(p)}$ for $p=1,2,\cdots, n-1$ may be obtained from 
(\ref{IV10a}) and (\ref{IV11}) in a similar way like that of (\ref {IV16a}). 
The integral equations for the kernels ${\bf K}^{(p)}$ thus turn out to be of 
the form
\begin{equation}
{\bf K}^{(p)}(x,z) - e_n F_p^*(x+z) - \int _x^{\infty}dy
{\bf K}^{(n)}(x,z)F_p^*(z+y) =0
\label{IV16b}
\end{equation}
\label{IV16}
\end{subequations}
provided $z>x$. In deriving (\ref {IV16b}) we have used the following identity 
\[
C_{np}^* = \alpha_{np}^*(\lambda_j)= 
\sum_{i=1}^{n-1}[{\tilde \alpha_{np}}(\lambda_j)]_{ki} 
(Adj[{\tilde \alpha_{np}}(\lambda_j)])_{il} 
\delta_{kl}.
\]
The set of coupled equations (\ref {IV16}) may be called as generalized Gelfand 
Levitan Marchenko equations. Substituting now (\ref {IV12b}) and (\ref{IV16b}) 
in (\ref{IV16a}), to a first approximation, we find the Gelfand Levitan 
Marchenko equation for the $p$th. component of ${\bf K}^{(n)}$:   
\begin{equation}
K^{(n)}_p(x,z) +F_p(x+z) +\sum_{m=1}^{n-1} \int_x^{\infty}ds K^{(n)}_p(x,s)
\int_x^{\infty}dyF_m(y+z)F^*_m(y+s) =0
\label{IV17}
\end{equation}
which will be used later to find the soliton solutions for VHNLS equataion. 

\section{\bf N Soliton Solutions}

To obtain soliton solutions, let us associate dynamical fields $q_i(x,t)$ with 
the kernels, $K^{(n)}_i$ in (\ref {IV17}). 
Substituting (\ref{IV11b}) into Lax equation (\ref{2.1}a) we find 
\begin{equation}
q_i(x)= -2K^{(n)}_i (x,x)
\label{V1}
\end{equation}
and consequently, it is evident that the solution of (\ref {IV17}) gives rise to 
soliton solutions in terms of scattering data elements. But before going to 
solve (\ref {IV17}), we 
first compute time evolution of scattering data element. Notice that as 
$\vert x \vert \rightarrow \infty$, the Lax equation (\ref {2.2b},\ref {2.3}b) 
leads to
\begin{equation}
\frac{\partial \Psi}{\partial t} =-4i\epsilon \lambda^3\Sigma
\label{V2}
\end{equation}
which, in turn, determines time evolution of the scattering data elements.
For example, scattering data elements, $\alpha_{nj}(\lambda,t)$ for $j=1,2,
\cdots, n-1$ evolve with time as
\begin{subequations}
\begin{equation}
\alpha_{nj}(\lambda,t) = \alpha_{nj}(\lambda,0)e^{-8i\epsilon \lambda^3t}
\label{V3a}
\end{equation}
while the element $\alpha_{nn}(\lambda,t)$ is time invariant:
\begin{equation}
\alpha_{nn}(\lambda,t) = \alpha_{nn}(\lambda,0),
\label{V3b}
\end{equation}
\label{V3}
\end{subequations}
which eventually justifies our conjecture in section 3 that $\alpha_{nn}$, 
indeed, can be associated with the conserved quantities. From (\ref {IV8}) and
(\ref {IV14b}) it follows that the coefficients $C^{(j)}_{np}$ also satisfy the
similar time dependence as $\alpha_{nj}$ in (\ref {V3a}) and thus
\begin{equation}
C^{(j)}_{np}(\lambda,t) = C_{np}^{(j)}(\lambda,0)e^{-8i\epsilon 
\lambda^3 t}
\label{V4}
\end{equation}

If we now restrict ourselves to soliton sector, $\alpha_{np}(\lambda)$ becomes 
trivial and the function $F_p(x+y)$, in this case, reduces to   
\[
F_p(x+y)=i\sum_{j=1}^{N}\frac{C_{np}^{(j)}e^{-i\lambda^*_j(x+y)}}
{\alpha_{nn}^{\prime}(\lambda^*_j)}. 
\]
Time dependence of the function $F_p(x+y)$ may be obtained immediately from 
(\ref {V4}) as
\begin{equation}
F_p(x+y)=i\sum_{j=1}^{N}\frac{C_{np}^{(j)}(\lambda,0)e^{-8i\epsilon \lambda^{*3}_jt}
e^{-i\lambda^*_j(x+y)}}{\alpha_{nn}^{\prime}(\lambda^*_j)}.
\label{V6}
\end{equation} 

To solve the integral equation (\ref {IV17}) for the soliton solutions, the 
kernels $K^{(n)}_p(x+y)$ are assumed to be of the form 
\begin{equation}
K_p^{(n)}(x+y) = \sum_{j=1}^{N}\omega_{pj}(x,t)e^{-i\lambda^*_j y}
\label{V7}
\end{equation}
Substituting (\ref {V6}) and (\ref {V7}) in (\ref {IV17}), we obtain a set of
$n-1$ algebraic equations:
\begin{eqnarray}
K^{(n)}_p(x,x)&+&F_p(x+x)-\sum_{j,k,l=1}^{N}\sum_{r=1}^{n-1}\frac{\omega_{pl}
e^{-i\lambda_j^*x}}{(\lambda_k-\lambda_j^*)(\lambda_k-\lambda_l^*)}\cdot
\nonumber\\
&~&\frac{C^{(j)}_{nr}(0)C^{*(k)}_{nr}(0)}{\alpha^{\prime}_{nn}(\lambda_j^*)
\alpha^{*\prime}_{nn}(\lambda_k)}
e^{i(2\lambda_k-\lambda_j^*-\lambda_l^*)x+8i\epsilon (\lambda_k^3-\lambda_j^{*3})t}=0
\label{V8}
\end{eqnarray}
Solving (\ref {V8}) for $K_p(x,x)$ and subsequently using (\ref {V1}), the $N$ 
soliton solutions for each dynamical field $q_i(x,t)$ may be expressed as
\begin{equation}
q_i(x,t) = -2\sum^{N}_{j=1}({\bf B}{\bf C}^{-1})_{ij}e^{-i\lambda_j^*x}
\label{V9}
\end{equation}
where, ${\bf B}$ and ${\bf C}$ are respectively $(n-1)\times N$ and $N\times N$
matrices whose explicit forms are given by 
\[
({\bf B})_{ij} = i C^{(j)}_{ni}(0)e^{-8i\epsilon\lambda_j^{*3}t - 
i \lambda_j^{*}x}
\]
\[
({\bf C})_{ij} = \sum_{p=1}^{n-1}\sum_{k=1}^N \frac{C^{(j)}_{np}(0) 
C^{(k)}_{np}(0)e^{-i(\lambda_i^*+\lambda_j^*-2\lambda_k )x 
+ 8i\epsilon(\lambda_k^3-\lambda_j^{*3})t}}{\alpha^{\prime}_{nn}(\lambda_j^*)
\alpha^{*\prime}_{nn}(\lambda_k)(\lambda_k-\lambda_j^*)(\lambda_k-\lambda_i^*)} 
-\delta_{ij}
\]

Let us now consider an explicit form of one soliton. One soliton solution, 
which follow from (\ref {V9}), may be written in the following form
\begin{equation}
q_i(x,t) = \frac{2F_i(x+x)}{1 + \sum_{p=1}^{n-1}
          \vert F_p(x+x) \vert^2 \frac{1}{(\lambda_1 -\lambda_1^*)^2}}
\label{V10}
\end{equation}
where
\[ 
F_i(x+y) = i\frac{C_{ni}^{(1)}(\lambda;0)}
{\alpha_{nn}^{\prime}(\lambda^*_1)}e^{-8i\epsilon \lambda^{*3}_{1}t 
- i \lambda^*_1(x+y)}.
\]
$q_i(x,t)$ in (\ref {V10}) may be expressed in a more conventional way, by 
choosing the position of the pole at $\lambda_1=\frac{1}{2}(-\xi+i\eta)$ and 
by introducing $e^{\gamma_i+i\delta_i}=\frac{C_{ni}^{(1)}(\lambda,0)}
{2\eta \alpha_{nn}^{\prime}(\lambda^*_1)}$ and thus
\begin{equation}
q_i(x,t) = 4i\eta \frac{e^{\gamma_i-\eta x+\epsilon\eta(\eta^2-3\xi^2)t
+i(\delta_i+\xi x+\epsilon\xi^3t-3\epsilon\xi^2\eta t)}}{1+\sum_{p=1}^{n-1}
e^{2\gamma_p}e^{-2\eta x+2\epsilon(\eta^3-3\eta\xi^2)t}}
\label{V11} 
\end{equation}
which can further be simplified as
\begin{equation}
q_i(x,t) = \frac{2i\eta e^{\gamma_i-\Gamma_n + iQ_i}}{coshP}
\label{V11a}
\end{equation}
by introducing 
\begin{eqnarray*} 
e^{2\Gamma_n} &=& \sum_{p=1}^{n-1}e^{2\gamma_p}\\ 
P(x,t) &=& \eta x-\epsilon \eta(\eta^2-3\xi^2)t-\Gamma_n\\
Q_i(x,t)&=&\xi x+\epsilon \xi(\xi^2-3\xi\eta)t+\delta_i\\
\end{eqnarray*} 

Once again, if we specialize to HNLS equation, each independent field $q_i(x,t)$, 
depending on the models, reduces either to $q$ or to $q^*$ and as a consequence 
$q_i$ in (\ref {V11}) yields to
\begin{equation}
q(x,t) = \frac{2i\eta e^{i{\tilde B}}}{\sqrt{(n-1)} cosh{\tilde A}}
\label{V12}
\end{equation}
where, ${\tilde A}= \eta x-\epsilon\eta(\eta^2-3\xi^2)t-\gamma
-\frac{1}{2}ln(n-1)$ and ${\tilde B}=\xi x+\epsilon\xi(\xi^2-3\xi\eta)t+\delta$. 
It is interesting to note that we have obtained precisely the same expression 
for one soliton in \cite {11}.  
\vspace{.5 cm}

\section{\bf Generalized Landau Lifshitz type equation as the Gauge 
equivalence system}  

We now show an interesting connection between the VHNLS equation 
and the generalized Landau Lifshitz type equation by exploiting the gauge
equivalence of the Lax pairs of these two dynamical systems. The procedure is 
similar to that between the nonlinear Schr{\" o}dinger equation and the standard 
Landau Lifshitz equation \cite {14}.

Under a local gauge transformation, the Jost function, ${\bf \Psi}(x,t,\lambda)$ 
changes to 
\begin{equation}
\tilde{{\bf \Psi}} = g^{-1}(x,t){\bf \Psi}(x,t,\lambda)
\label{III1}
\end{equation}
where $g(x,t) = {\bf \Psi}(x,t,\lambda)\vert_{\lambda =0}$. We claim that 
$g(x,t)$ is an element of $SU(n)$ group. As a consequence of the gauge 
transformation (\ref {III1}), the Lax equations (\ref {2.2},\ref{2.3}) become 
\begin{subequations}
\begin{eqnarray}
\tilde{{\bf \Psi_x}} &=&  \tilde{{\bf U}}(x,t,\lambda)\tilde{{\bf \Psi}}\\
\tilde{{\bf \Psi_t}} &=&  \tilde{{\bf V}}(x,t,\lambda)\tilde{{\bf \Psi}}
\end{eqnarray}
\label{III2}
\end{subequations}
where $\tilde{U}$ and $\tilde{V}$are the new gauge transformed 
Lax pair, given by 
\begin{subequations}
\begin{eqnarray}
 \tilde{{\bf U}}(x,t,\lambda) &=& g^{-1}({\bf U}-{\bf U}_0)g\\ 
 \tilde{{\bf V}}(x,t,\lambda) &=& g^{-1}({\bf V}-{\bf V}_0)g 
\end{eqnarray}
\label{III3}
\end{subequations}
with ${\bf U}_0 ={\bf U}\vert_{\lambda=0}=g_x(x,t)g^{-1}(x,t)$ and 
${\bf V}_0 ={\bf V}\vert_{\lambda=0}=g_t(x,t)g^{-1}(x,t)$. This leads to  
\begin{equation}
{\bf A}=g_x(x,t)g^{-1}
\nonumber
\end{equation}
Since ${\bf A}$ belongs to $su(n)$ algebra, $g(x,t)$ obviously belongs to 
$SU(n)$ group, which justifies our claim. 
We may now identify the spin field of the Landau Lifshitz type equation as
\begin{equation}
S =g^{-1}(x,t){\bf \Sigma}g(x,t), \quad\quad\quad S^2 =1
\label{III4}
\end{equation} 
With this identification, the gauge transformed Lax pair (\ref{III1}) 
may be expressed in terms of the spin field $S$ (\ref{III4}) and its 
derivatives only, yielding
\begin{subequations}
\begin{eqnarray}
 \tilde{{\bf U}} &=&-i\lambda S\\
\tilde{{\bf V}} &=& -4i\epsilon \lambda^3S + 2\epsilon\lambda^2SS_x 
           +i\epsilon\lambda(S_{xx} + \frac{3}{2}SS_x^2)
\end{eqnarray}
\label{III5}
\end{subequations}
In deriving (\ref {III5}) we have used the following important identities 
\begin{eqnarray*}
SS_x &=& 2g^{-1}{\bf A}g\\ 
SS_x^2 &=& -4g^{-1}{\bf \Sigma}{\bf A}^2g \\
S_{xx} &+& SS_x^2 = 2 g^{-1}{\bf \Sigma}{\bf A}_xg
\end{eqnarray*}
The zero curvature condition of (\ref {III5}), namely 
\[
\tilde{{\bf U}}_t - \tilde{{\bf V}}_x + [\tilde{{\bf U}}, \tilde{{\bf V}}]=0
\]   
ultimately leads to the generalized Landau Lifshitz type equation
\begin{equation}
S_t+\epsilon S_{xxx}+\frac{3}{2}\epsilon (S_x^3+SS_{xx}S_x+SS_xS_{xx}) = 0
\label{III6}
\end{equation}
with$S\in SU(n)/(U(n-1))^n$.
\vspace{.5 cm}

\section{\bf Conclusion}

We have formulated a generalized IST method for an $n$ dimensional Lax oparator. 
Subsequently, the generalized IST method is used to obtain $N$ soliton solutions 
for a multi-component generalization of HNLS equation, proposed by us. We have, 
although, obtained soliton solutions for VHNLS equation,
 IST method developed here is quite general and 
is applicable for obtaining soliton solutions for all integrable nonlinear 
equations. 
The $sech$ structure of one soliton solution of 
VHNLS equation is quite interesting, particularly in the context of nonlinear optics, since it can easily be produced from the 
output of a mode locked laser. 

We have shown the integrability of VHNLS equation also in the Liouville sense. 
This has been acheived first by finding a set of coupled Riccati equations and 
subsequently by identifying the generating function
for the conserved charges. It is found that the last diagonal element of 
the scattering data matrix may be identified as the generating function of the 
conserved charges. This is also confirmed from the time evolution of scattering 
data matrix elements. We have also established an intriguing 
relationship between 
the VHNLS equation and a generalized Landau Lifshitz equation, where the spin
field $S\in SU(n)/(U(n-1))^n$. Moreover, we have obtained a Lax pair for the 
spin system implying direct integrability of the generalised Landau Lifsitz 
equation.

\vspace{1.0 cm}

S.N. would like to thank CSIR, Govt. of India for financial support and for 
the award of Junior Research Fellowship.

Electronic address : $^*$sasanka@iitg.ernet.in, $^{\dag}$sudipta@iitg.ernet.in

\end{document}